\documentclass{aastex6}
\usepackage{subfigure}
\usepackage{natbib}
\usepackage{graphicx}
\usepackage{amssymb}  
\usepackage{amsmath}

\begin{document}
    \title{Bright hot impacts by erupted fragments falling back on the Sun: magnetic channelling}
 
\author{A.Petralia\altaffilmark{1,2}} 
\and
\author{F.Reale\altaffilmark{1,2}}
\and
\author{S.Orlando\altaffilmark{2}}
\and
\author{P.Testa\altaffilmark{3}}

\altaffiltext{1}{Dipartimento di Fisica \& Chimica, Universit\`a di Palermo,
              Piazza del Parlamento 1, 90134 Palermo, Italy}
\altaffiltext{2}{INAF-Osservatorio Astronomico di Palermo, Piazza del Parlamento 1, 90134 Palermo, Italy}
\altaffiltext{3}{Harvard-Smithsonian Center for Astrophysics, Cambridge, MA 02138, USA}

   \date{Received ; accepted }
   \begin{abstract}
Dense plasma fragments were observed to fall back on the solar surface by the Solar Dynamics Observatory after an eruption on 7 June 2011, producing strong EUV brightenings. Previous studies investigated impacts in regions of weak magnetic field. Here we model the $\sim~300$ km/s impact of fragments channelled by the magnetic field close to active regions. In the observations, the magnetic channel brightens before the fragment impact. We use a 3D-MHD model of spherical blobs downfalling in a magnetized atmosphere. The blob parameters are constrained from the observation. We run numerical simulations with different ambient density and magnetic field intensity. We compare the model emission in the 171\AA~ channel of the Atmospheric Imaging Assembly with the observed one. We find that a model of downfall channelled in a $\sim~1$MK coronal loop confined by a magnetic field of $\sim~10-20$G, best explains qualitatively and quantitatively the observed evolution. The blobs are highly deformed, further fragmented, when the ram pressure becomes comparable to the local magnetic pressure and they are deviated to be channelled by the field, because of the differential stress applied by the perturbed magnetic field. Ahead of them, in the relatively dense coronal medium, shock fronts propagate, heat and brighten the channel between the cold falling plasma and the solar surface. This study shows a new mechanism which brightens downflows channelled by the magnetic field, such as in accreting young stars, and also works as a probe of the ambient atmosphere, providing information about the local plasma density and magnetic field.
\end{abstract}

\section{Introduction}

In the process of accretion in young stars, dense plasma flows through magnetic channels, which link the star with its disk, and impacts the stellar surface \citep{Uchietal1984,Bertetal1988}. An emission excess and associated high plasma densities have been observed in high energy bands, and have been related to the accretion process \citep{Kastetal2002,Tellesetal2007,Argiretal2007,Argiretal2011,Testetal2004}. The formation of shocks after the impact could explain this excess \citep{Orletal2010} and also the complexity of the magnetic field in the impact region can influence the process. Impacts of dense plasma have been observed also in the solar corona with local brightenings in the EUV band that recall the emission excess in young stars. On the Sun we can study the process in great detail (e.g., \citealt{Xiaetal2016}) and we can characterize the role of each player, e.g., magnetic field, downfalling plasma, and shocks.

In a solar eruption triggered by a flare on 7 June 2011, a large amount of fragments of an original dense and cold filament fall back spreading far from the original place all around on the solar surface \citep{Carletal2014,Doletal2014,Drietal2014,Innetal2012,Reaetal2013,Reaetal2014}. Part of the dense fragments falls in quiet Sun where the magnetic field was weak. Their impacts on the solar surface were bright in the EUV band observed with SDO/AIA. These brightenings are consistently reproduced by hydrodynamic modelling and recall the excess of high energy emission observed in accreting young stars \citep{Reaetal2013}. The fragments we analysed in \cite{Reaetal2013} were shown to travel along ballistic trajectories, and, therefore, the magnetic field had a minor role throughout their evolution.

Many other fragments however appear to fall in regions where the magnetic field is much stronger, and even inside active regions. These fragments show a different evolution and destiny. In particular, we no longer see bright impacts but the fragments are deviated, channelled and the whole final segment of the channels is activated into bright thinner filaments.

In this work we address this different class of downfalling fragments. It is clear from the observations that here the magnetic field plays a different and critical role in determining the evolution of the blobs and, thus, the mechanism that produces the excess of the emission is not necessary the same as that indicated in the previous work. So it is interesting to explore these cases in which the interaction of the blobs with the magnetic field is important.

Our approach is similar to that of \cite{Reaetal2013} though here we need to include the description of an appropriate ambient magnetic field and therefore a full magnetohydrodynamic model. The fragments do not follow a simple trajectory but they are deviated as they move deeper and deeper in the low corona, confirming a non-trivial interaction with the more and more intense magnetic field. Another different and fundamental ingredient is that the downfalling fragments are eventually forced to propagate inside an already dense and hot medium, that is the plasma confined inside active region loops. This plasma will be strongly perturbed and activated by the infalling material, which will then act also as a probe for the ambient corona.
 
This case represents a unique opportunity to probe active region conditions and their reaction to strong perturbations coming from outside. On the other hand, this is also closer to the conditions in star forming regions, where the flows coming from the circumstellar disk are believed to be funneled by the magnetic channels that link the disk to the young stars. 

In summary, this is an excellent opportunity to study the funneling of downfalling plasma and its interaction with the possible dense corona close to the stellar surface.
The observed phenomenon is presented in Section 2, the model is described in Section 3, the simulations and the results in Section 4, and we discuss them in Section 5.

\section{The observation}
\label{sec:observ}

In this work, we study the interaction of downfalling fragments of plasma, spread onto the solar surface by a solar eruption after an M-class flare in 7 June 2011. The event was observed by the Atmospheric Imaging Assembly (AIA) \citep{Lemenetal2002} on board the Solar Dynamics Observatory (SDO) \citep{Pesnetal2012}, in the ultraviolet (UV) and extreme-ultraviolet (EUV) narrow-band channels. The AIA instrument provides data with a high cadence ($\sim 12$ s) and high spatial resolution ($\sim 0.6$ arsec per pix).

In this event, a dense and cold filament is broken into many fragments erupted in all directions. Part of them fall back onto the solar surface, far from the eruption location. 

\begin{figure*}[!t]
    \centering
    \includegraphics[width=10cm]{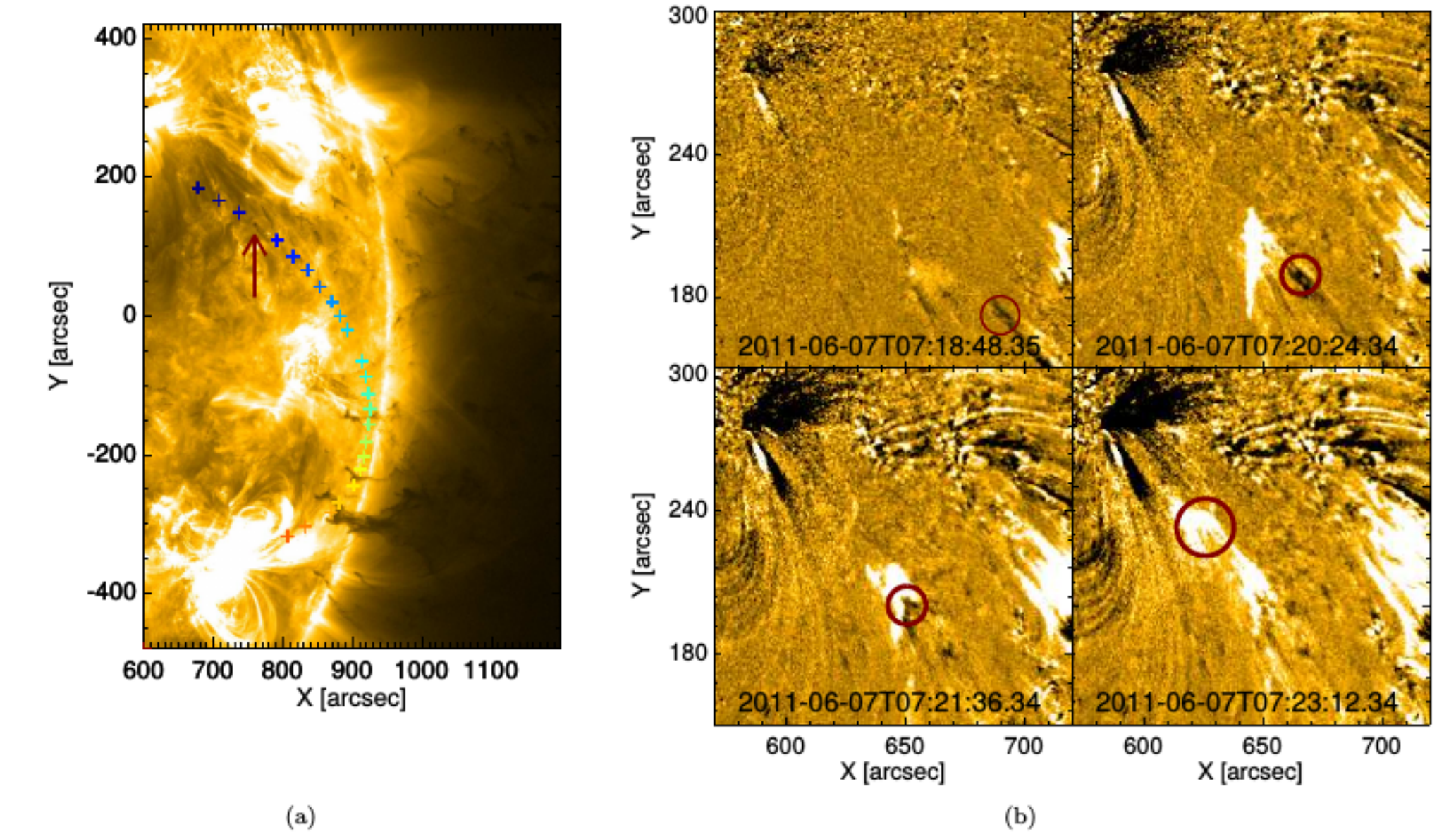}
\caption{(a) Time increases from red to blue (crosses are spaced by $\sim 120$ s): Path followed by a fragment of the erupted plasma from the flaring active region  to the impact active region. Time progresses from red crosses (6:30 UT) to blue crosses (7:19 UT). The background image is taken in the AIA 171 \AA~ channel at time 7:13 UT (the position of the fragment at this time is pointed by the arrow). (b) Images (subtracted by the one at 7:18 UT) of the final evolution of the fragment in (a), marked by red circles, impacting the solar surface in the proximity of the active region at 4 subsequent times. See Supplementary Movie 1.}
 \label{fig:iniobs}
 \end{figure*}

\begin{figure*}[!t]
	\centering
	\includegraphics[width=0.4\columnwidth]{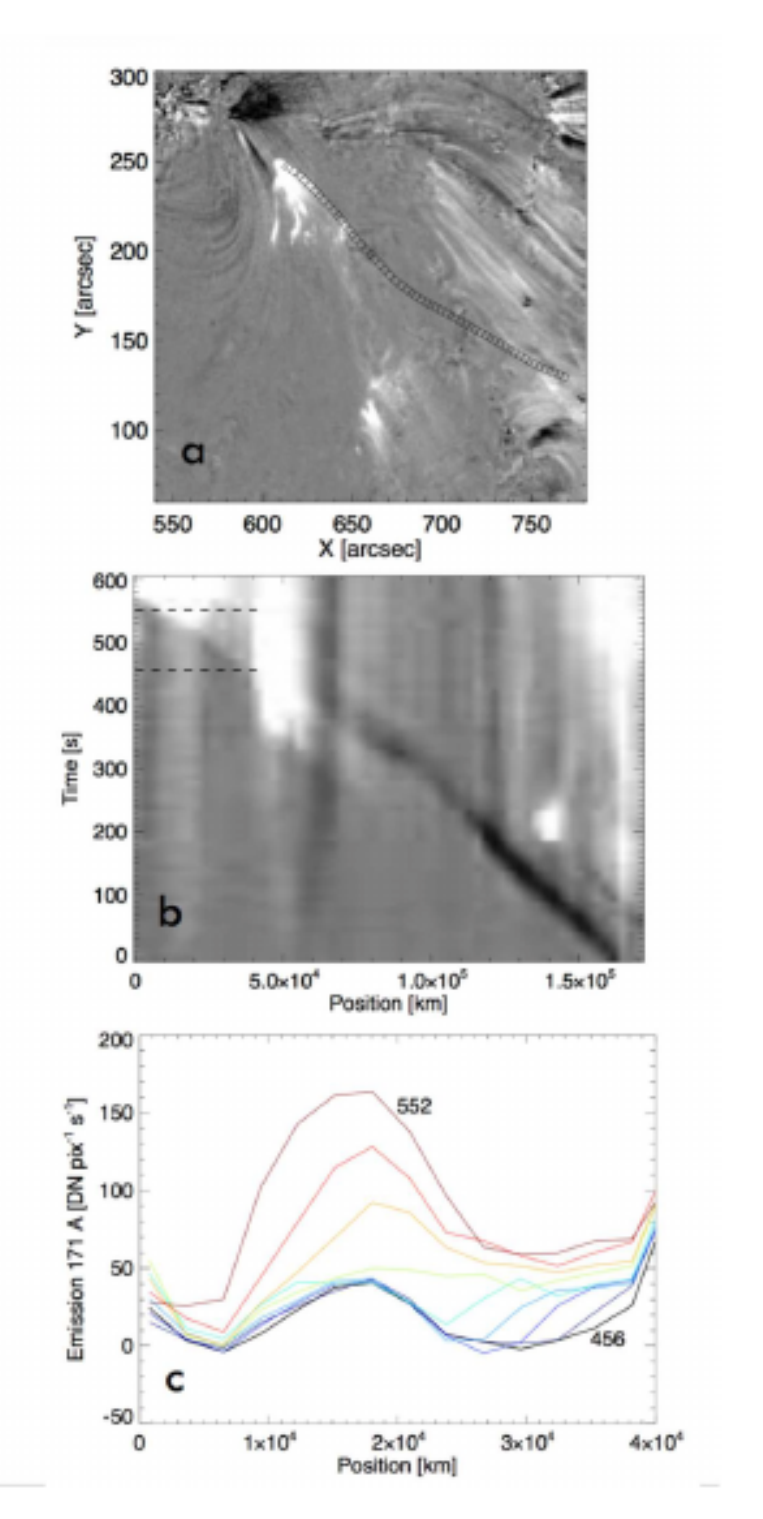}
	\caption{171 \AA~ difference image at 7:24 UT, after subtracting the first in the sequence (7:10 UT), in which we track the position of the falling blob. A clear change in the trajectory can be seen at $\sim (680,180)$ arcsec, approaching the active region. (b) 171 \AA~ emission along the strip in panel (a) as a function of time, between 7:14 UT and 7:24 UT. The grey scale is in the range [-100,100] DN pix$^{-1}$ s$^{-1}$. The dashed line bounds the time and space range of panel (c). (c) 171 \AA~ emission along the strip in panel (a) zoomed in the space and time range between the dashed lines in panel (b). The lines are spaced by 12 s, with the time increasing from blue to red. The initial and final curves are at the labelled times (s).}
	\label{fig:blobtraj1}
	\end{figure*}

Some fragments fall outside of active regions, where the magnetic field is weak, so they show ballistic trajectories until they hit the solar surface. In the proximity of the impact region, a brightening is observed \citep{Reaetal2013}. Other fragments instead fall close or inside active regions, and here we focus on one of these active regions (the one whose centre is [580,280] arcsec from the disk center). Fig.~\ref{fig:iniobs}a shows the entire trajectory of one of these fragments \citep[tracked with an automatic detector of local emission minima,][]{Reaetal2013}. The fragment follows a ballistic motion as long as it is far from the active region, but close to it, it is deviated, as clear from the final part of the path shown in Fig.~\ref{fig:blobtraj1}a. Fig.~\ref{fig:iniobs}b shows images at 3 subsequent times of the fragment final evolution in the AIA 171 \AA~ channel. A brightening is observed already as it is being channelled by the magnetic field, before impacting the solar surface. This bright front precedes this fragment (and others) and propagates ahead of it along the entire magnetic flux tube. The fragment disappears once it is completely channelled.
During the fall, before they are channelled, the fragments are dark in all the AIA EUV channels. They change shape but remain small compact blobs.

Since the blobs fall ballistically until the interaction with the magnetic field causes a change in the trajectory, their velocity component perpendicular to the solar surface can be estimated by the simple formula: 

\begin{equation}
 v_{ff} = \sqrt{2 g_{\odot} D} \sim 240 ~ km/s
 \label{eq:free-fall}
\end{equation}

\noindent
where $g_{\odot} \sim 2.74 \times 10^4$ cm s$^{-2}$ is solar gravity and $D \sim 10^{10}$ cm is the maximum height reached by the blobs above the surface. This height has been estimated as the distance of the apex from the line connecting the footpoints. Therefore we are assuming that, since the event is close to the solar limb, we are seeing the trajectory in Fig.~\ref{fig:iniobs}a face-on and with no other tilting. Moderate differences from this assumption do not lead to substantial variations of $v_{ff}$ (for a tilted trajectory of $\sim 30^o$ the error is $\sim 10 \%$). 
To estimate the velocity component parallel to the solar surface we make the rough assumption that in the very final part of the trajectory (Fig.~\ref{fig:iniobs}b) we detect mostly the motion projected on the solar surface. We measure a projected length of $\sim 4 \times 10^{9}$ cm that is covered in a time of $\sim 200$ s, which corresponds to a horizontal speed of $v_h \sim 200$ km/s.  
The total velocity of the blobs is then estimated to be $v_{tot} = \sqrt{v_{ff}^2 + v_h^2} \sim 300$ km/s.
 
Along the final part of the trajectory, we have also defined a strip and divided it into approximately square sectors. In each sector we have evaluated the average emission and subtracted the value at an early time (7:10 UT). Fig.~\ref{fig:blobtraj1}b shows the resulting emission profiles along the path as a function of time ($600$~s, from 7:14~UT to 7:24~UT). The blob is dark at early times, when it is still far from the active region. After $t \sim 200$ s it begins to fade away and for $t > 400$ s it turns into an bright feature. The bright features at position $\sim 5 \times 10^4$ km ($t > 300$ s) and $1.4 \times 10^{5}$ km ($t \sim 200$ s) are not moving ones. Fig.~\ref{fig:blobtraj1}c show profiles of the emission    
zoomed close to the active region and in the final stage of the evolution. The fronts move from right to left and become increasingly bright, with an emission rate in the range $50-150$ DN/s/pix.

Before brightening, the blobs are dark. From the amount of absorption one can estimate their density as described in \cite{LaneRea2013}. We obtain a density ranging between $1$ and $4 \times 10^{10}$ cm$^{-3}$. For our simulation, we assume $2\times10^{10}$ cm$^{-3}$ as the blob density of our reference case, but in other cases we consider a blob density of $10^{10}$ cm$^{-3}$ (see section \ref{sec:themodel}).

\section{MHD Modelling}

\subsection{The model}
\label{sec:themodel}

Our model solves the magnetohydrodynamic (MHD) equations for an ideal compressible plasma in the following conservative form:

        \begin{equation}
        \dfrac{\partial \rho}{\partial t} + \triangledown \cdot (\rho \textbf{v}) = 0
        \end{equation}
        \begin{equation}
        \dfrac{\partial \rho \textbf{v}}{\partial t} + \triangledown \cdot (\rho \textbf{v} \textbf{v} - \textbf{B}\textbf{B} +p_{t}\textbf{I}) = \rho  \textbf{g}
        \end{equation}
        \begin{equation}
        \dfrac{\partial E}{\partial t} + \triangledown \cdot ( (E+p_{t} ) \textbf{v} - \textbf{B}( \textbf{v} \cdot \textbf{B}) ) = \rho  \textbf{v}  \cdot \textbf{g} - n_{e}n_{H} \Lambda(T) + H - \triangledown \cdot \textbf{F}_{c}
        \end{equation}
        \begin{equation}
        \dfrac{\partial  \textbf{B}}{\partial t} + \triangledown  \cdot ( \textbf{v} \textbf{B} - \textbf{B} \textbf{v} ) = 0
        \end{equation}
        \begin{equation}
        \triangledown \cdot \textbf{B} = 0
        ,\end{equation}
where
        \begin{equation}
        \rho = \mu m_{H} n_{H}
        \end{equation}
        \begin{equation}
        p_{t} = p +  \frac{ \textbf{B} \cdot \textbf{B}}{2}
        \end{equation}
        \begin{equation}
        E =\rho \epsilon+ \rho \frac{\textbf{v} \cdot \textbf{v}}{2} + \frac{ \textbf{B} \cdot \textbf{B}}{2}
        \end{equation}
        \begin{equation}
        \textbf{F}_{c} = \frac{ F_{sat}}{F_{sat} + |\textbf{F}_{class}|} \textbf{F}_{class}
        \label{eq:condflux}
        \end{equation}
        \begin{equation}
        \textbf{F}_{class} = k_{\parallel} \textbf{b} (\textbf{b} \cdot \bigtriangledown T ) + k_{\perp} ( \bigtriangledown T -
\textbf{b} (\textbf{b} \cdot \bigtriangledown T ) )
        \end{equation}
        \begin{equation}
        |\textbf{F}_{class}| = \sqrt{(\textbf{b} \cdot \bigtriangledown T)^{2}(k_{\parallel}^{2}-k_{\perp}^{2}) + k_{\perp}^{2}\bigtriangledown T^{2}}
        \end{equation}
        \begin{equation}
        F_{sat} =5 \Phi \rho c_{s}^{3}
        \end{equation}

\noindent        
where $\rho$ is the density per unit mass, $\mu = 1.265$ is the mean atomic mass (assuming solar metal abundances; \citealt{AndGrev89}), $m_{H}$ is the mass of the hydrogen atom, $n_{e}$ and $n_{H}$ are, respectively, the electron and hydrogen number density, $p_{t}$ is the total pressure, that is, the sum of the thermal pressure $p$ and the magnetic pressure (the factor $ 1/ \sqrt{4 \pi}$ is absorbed in the definition of the magnetic field $\textbf{B}$), $E$ is the total energy density, that is, the sum of the thermal energy density ($\rho \epsilon$), the kinetic energy density and the magnetic energy, $\textbf{v}$ is the plasma velocity, $\textbf{g}=g_\odot \hat{\textbf{z}}$ is the solar gravity, $\hat{\textbf{z}}$ is the unit vector along the vertical direction, $\textbf{I}$ is the identity tensor, $T$ is the temperature, $\Lambda(T)$ is the radiative loss function for optically thin plasma, $ \textbf{F}_{c}$ is an anisotropic (i.e. along the magnetic field lines) flux-limited expression that varies between the classical and saturated thermal conduction regimes $\textbf{F}_{class}$ and $F_{sat}$ respectively, $k_{\parallel} = K_{\parallel}T^{5/2}$ and  $k_{\perp}=K_{\perp}\rho^2/(B^2T^{1/2})$ are thermal conduction coefficients along and across the magnetic field,  $K_{\parallel}$ and $K_{\perp}$ are constants, $\textbf{b}$ is the magnetic unit vector, $H$ is a heating function whose only role is to keep the unperturbed atmosphere in energy equilibrium, $c_{s}$ is the sound speed for an isothermal plasma, $\Phi$ is a free parameter ($<1$, \citealt{Giul84}) that determines the degree of saturation of the thermal conduction; we set $\Phi = 0.9, $  which corresponds to quite an efficient conduction, more typical of coronal conditions \citep{Cowetal1977}. For the description of the flux-limited thermal conduction in Eq.~\ref{eq:condflux}, we adopted the same procedure for smoothly implementing the transition from the classical to saturated conduction regime introduced by \cite{Balbetal1982} (see also \citealt{Daltetal1993,Orletal2005}) which is a standard parametrization largely used in the literature. We completed this set with the equation of state for an ideal gas:
        \begin{equation}
        p = (\gamma -1)\rho \epsilon
        \end{equation}
\noindent
where $\gamma = 5/3$ is the adiabatic index.        
        
The calculations have been performed using PLUTO \citep{MignBod07,MignZan12}, a modular, Godunov-type code for astrophysical plasmas. The code provides a multiphysics, algorithmic modular environment particularly oriented toward the treatment of astrophysical flows in the presence of discontinuities, as in the case treated here. The code was designed to make efficient use of massive parallel computers using the message-passing interface (MPI) library for interprocessor communications. The MHD equations have been solved using the MHD module available in PLUTO, configured to compute inter-cell fluxes with the Harten-Lax-Van Leer approximate Riemann solver, while second-order in time was achieved using a Runge-Kutta scheme. A monotonized central difference limiter for the primitive variables have been used. The evolution of the magnetic field was carried out by adopting the constrained transport approach \citep{Bals99} that maintains the solenoidal condition ($\triangledown \cdot B = 0$) at machine accuracy. PLUTO includes optically thin radiative losses in a fractional step formalism \citep{MignBod07}, which preserves the second-order time accuracy because the advection and source steps are at least accurate to second order; the radiative losses $\Lambda(T)$ values have been computed at the temperature of interest using a table lookup/interpolation method using CHIANTI code (Version 7) \citep{LanZan12}, assuming a density of $10^9$ cm$^{-3}$ and ionization equilibrium according to \citet{Dere09}. We assume energetic equilibrium in the chromosphere and inside the initial cold blobs, therefore we set $\Lambda(T)= 0$, as well as $ H = 0 $, for $T < 10^4$ K. The thermal conduction was treated separately from advection terms through the super-time-stepping technique \citep{AlexVas96} that speeds up explicit time-stepping schemes for parabolic problems.

\subsection{Initial and boundary conditions}

We study the evolution of the downfalling fragments in the magnetic field with detailed modelling of template blobs. In particular, we describe the evolution of four downfalling blobs across a magnetized and relatively dense corona. The ambient magnetic field is not aligned to the initial direction of blobs downfall. Therefore the configuration has no special symmetry and we need a full 3D description. However, we can assume a symmetric magnetic field with respect to a plane perpendicular to the surface and crossing the middle of the domain and of the blobs. The blobs will not acquire average motion components in the horizontal direction across the magnetic field and therefore we will not need a large domain extension in that direction, which we assume to be the $Y$ direction.

To approach the configuration of a loop-populated active region but still keeping it manageable and simple, we consider a combination of magnetic dipoles, so that the magnetic field is symmetric with respect to the side boundaries and is closed in the low region close to the chromosphere.

The computational box is three-dimensional and cartesian (X,Y,Z) and extends over $4\times10^9$ cm in the X direction, $1.2\times10^9$ cm in the Y direction and  $6\times10^9$ cm in the Z direction. The Z direction is perpendicular to the solar surface. The mesh of the 3D domain is uniformly spaced along the three directions with $512\times128\times512$ cells, giving a cell size of $\sim80\times90\times120$ km. This provides a good compromise to have both good resolution in all directions (the domain is larger along $Z$) and reasonable computational times. The blobs are sufficiently well resolved (their diameter ranges 30-40 cells) and the resolution allows to have a steady initial atmosphere. 

In this box we consider an ambient relatively dense corona linked to a much denser chromosphere through a steep transition region. The corona is a hydrostatic atmosphere \citep{RTV1978} that extends vertically for $10^{10}$ cm. The chromosphere is hydrostatic and isothermal at $10^4$ K with a density at the base of $\sim 10^{16}$ cm$^{-3}$. The atmosphere is made plane-parallel along the vertical direction ($Z$).

Our simulation strategy is to freeze the parameters of the falling blobs (except for their density in one case), which are constrained from the observation, and to consider a few different conditions of the background atmosphere and magnetic field. In general, the topology of the magnetic field, combined with the atmosphere conditions, ensures that the blobs propagate in a medium in which the $\beta$ parameter of the plasma is highly varying, i.e. increasing while approaching the chromosphere.

We take as ``reference model'' (here after RM) the configuration in which the pressure of the background atmosphere ranges between 0.05 dyn cm$^{-2}$ at the top of the transition region and 0.01 dyn cm$^{-2}$ at $Z=10.5\times10^9$ cm. In this case, the ion density is $n_0 \sim 3.5\times10^7$ cm$^{-3}$ and the temperature is $T_0 \sim 1.1\times10^6$ K at $Z=10.5\times 10^9$ cm. The magnetic field intensity is $\sim 170$ G at the top of the transition region and $\sim~15$ G at the initial position of the blobs (see below).

\begin{figure*}[!t]
	\centering
	\includegraphics[width=12cm]{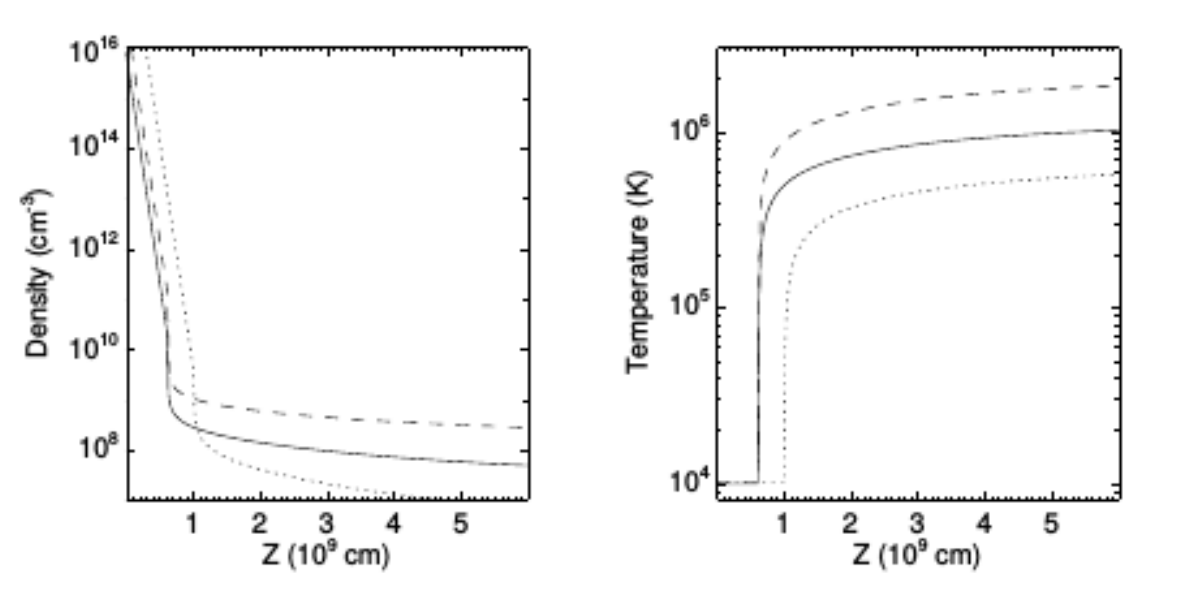}
	\caption{Density (left) and temperature (right) profiles of the reference (RM, solid line), cool (CM, dotted) and dense (DM, dashed) model atmospheres in the vertical direction ($Z$). }
	\label{fig:ini}
	\end{figure*}

We explore three other configurations that differ from the reference model either for the background atmosphere, or for the magnetic field intensity.
The second and third case, which are the 'dense model' (hereafter DM) and the 'cool model' (hereafter CM), differ from the RM for the hydrostatic conditions. The pressure ranges between 0.29 and 0.12 dyn cm$^{-2}$ (DM) and between 0.01 and 0.0006 dyn cm$^{-2}$ (CM), respectively. The ion density and the temperature are, respectively $\sim 2.2\times10^8$ cm$^{-3}$ and $\sim 2\times10^6 $K (DM) and $\sim 3.3\times10^6$ cm$^{-3}$ and $\sim 6.3\times10^5 $K (CM), at $Z=10.5\times 10^9$ cm. The fourth case is a 'weak field model' (here after WM) which differs from the reference model for the magnetic field intensity, which is about an order of magnitude lower, i.e., 1 G at the initial position of the blobs and $10$ G at the top of the transition region.
In all the models the position of the transition region varies between $Z=0.6\times10^9$cm and $Z=1\times10^9$cm. In Fig.\ref{fig:ini} we present the profiles of density and temperature of the ambient medium along the Z direction.

In all the cases, four blobs are put initially at a height in the range between $\Delta Z=3.5 \times10^9$ cm and $\Delta Z=4.5 \times10^9$ cm above the chromosphere, and at a distance in a range between $\Delta X=2.5 \times10^9$ cm and $\Delta X=4 \times10^9$ cm from the left boundary side, close to the upper right corner.
These length scales are in agreement with the path length that we measured in the observation (see section \ref{sec:observ}). For the sake of simplicity, we considered spherical blobs. Their radii are different, around the value we estimated from the data ($1.4-2\times10^8$ cm). The blobs have an initial downward vertical speed of $v = 300$ km/s and a temperature of $T = 10^4$ K. The initial temperature of the blobs is not very important because their evolution is much faster than any pressure readjustment with the ambient medium and we assume that initially they do not emit radiation. The blobs have a density of $2 \times 10^{10}$ cm$^{-3}$ in the RM, and $10^{10}$ cm$^{-3}$ in the other cases. The different density is not critical, but the choice provides the best match with the data for the reference case (RM, see Section~\ref{sec:s_emis}).

Boundary conditions are reflective at the left end of the X axis, the magnetic field is forced to be perpendicular to the boundary at the right end of the X axis, but for the other variables zero gradient has been set. Fixed conditions have been set at the lower end of the Z axis, and zero gradient at the upper end, with the exception of the magnetic field that is fixed. The same conditions are set at the far end of the Y axis. The computational domain is symmetric to a plane in Y=0, so we simulate half domain and set reflective conditions at the lower end of the Y axis.

\section{The simulations}
\label{sec:dynamics}

The reference model is our best model.
We now describe the evolution for the reference model and then we discuss separately how the other cases differ from this one.

\subsection{The Reference Model (RM)}

	\begin{figure*}[!t]
	\centering
	\includegraphics[width = 14cm]{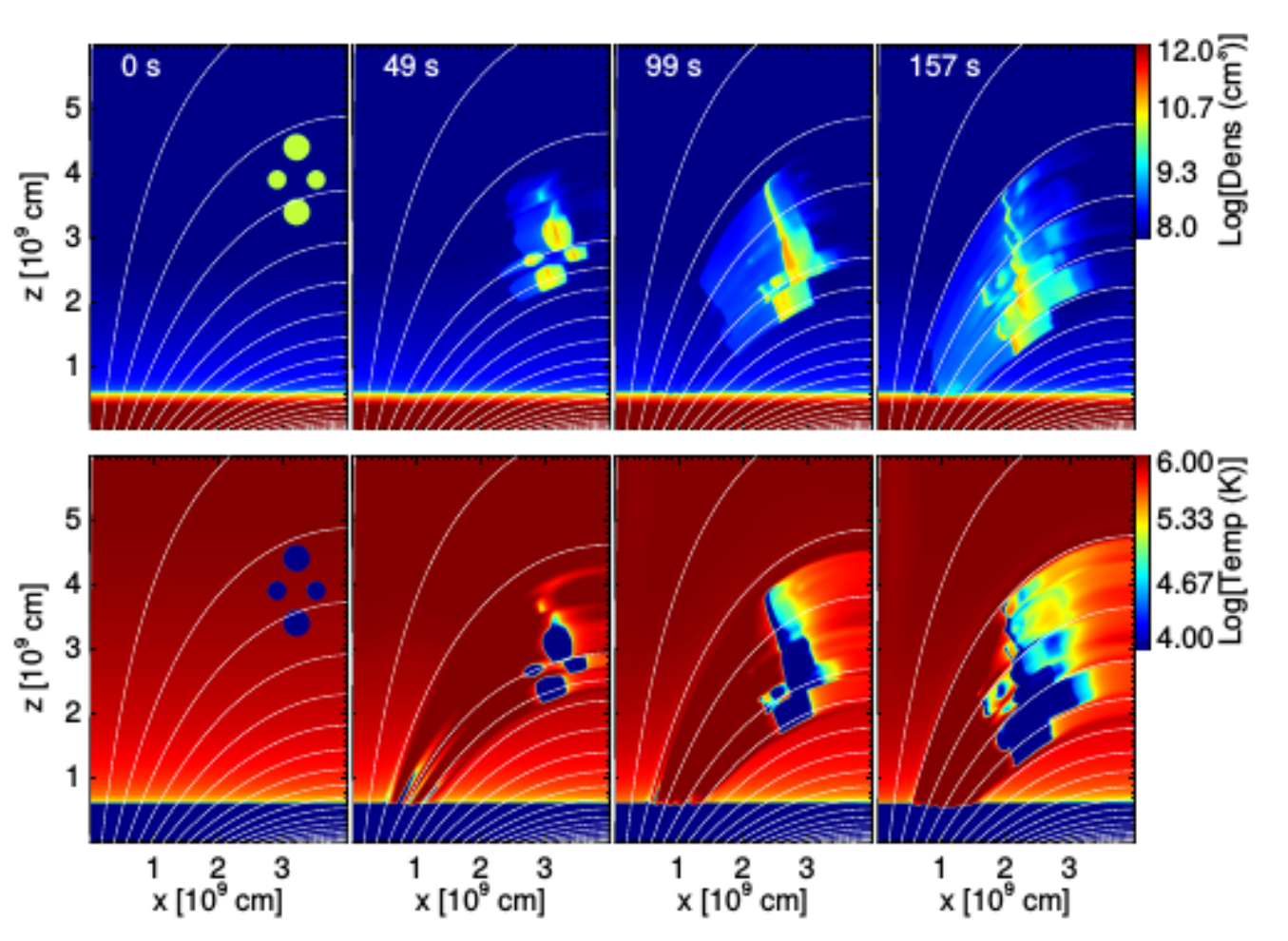}
	\caption{Reference simulation (RM): Density (top) and temperature (bottom) in the central cross-section X-Z of the domain, at four different times, in logarithmic scale. In all panels, magnetic field lines are shown. The color scales are saturated in the range of the palette. See also Supplementary Movie 2a-2b.}
	\label{fig:maps_rm}
	\end{figure*}

We present relevant snapshots of the density and temperature in a cross-section X-Z at the center of the domain, at the beginning and the 3 later times in Fig.~\ref{fig:maps_rm} (and Movie~2a-~2b). The blobs start to fall vertically by the gravity (see $t=49$ s), because their ram pressure ($p_r=\rho v^2$) largely  exceeds the surrounding magnetic pressure ($p_m=B^2/8\pi$) and the buoyancy is small:

\begin{equation}
p_r \sim  40 ~~ dyn ~ cm^{-2} >> p_m \sim 10 ~~ dyn ~ cm^{-2}
\label{eq:prescomp}
\end{equation}

During this vertical motion, that lasts for $\sim 60$~s, the blobs compress the magnetic field lines below them (as evident in the figure and in the movie\footnote{Some field lines are apparently not frozen to the plasma. This is an artifact of the 2D mapping of the field line reconstruction.}), and the magnetic pressure increases there. During this process, the magnetic field progressively brakes the blobs in the direction perpendicular to the field lines, until the magnetic pressure fully balances the ram pressure, and the blobs move only along the field lines. 

In this phase, the blobs are being compressed too, due to the braking by the magnetic field and to the interaction with the relatively dense ambient plasma, becoming slabs at higher density ($5\times 10^{10}$ cm$^{-3}$), but temperature still close to the initial value ($10^4K$). Behind these compressed blobs, wakes develop, with a coronal density but a  temperature not as high ($T \sim 6 \times 10^5K$). 

\begin{figure*}[!t]
	\centering
	\includegraphics[width=12cm]{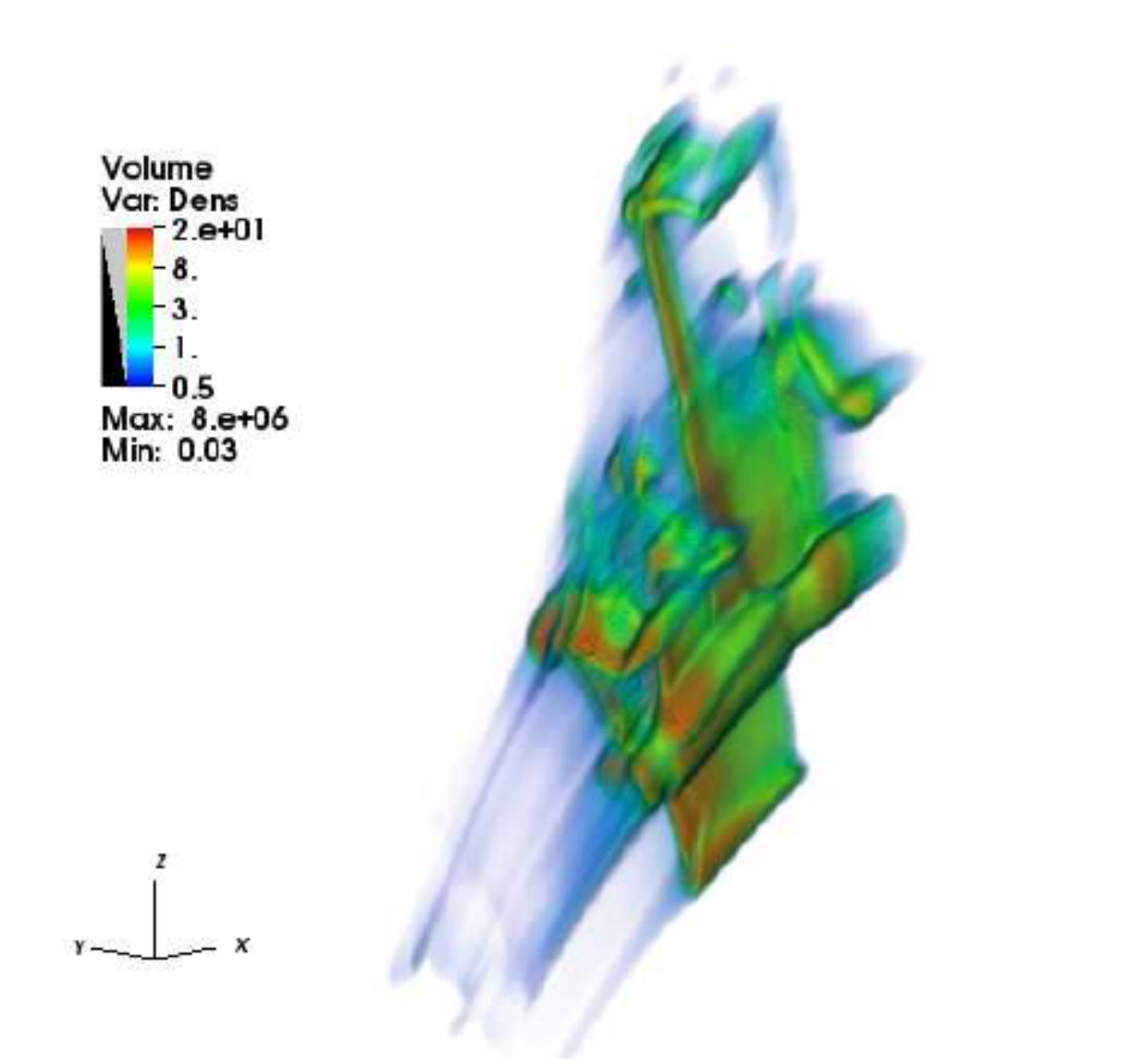}
	\caption{Rendering of the density (in units of $10^9$ cm$^{-3}$) of the blobs at $100$s. The blobs are denser (green-red) than the shocked plasma (blue). (See Supplementary Movie 3)}
	\label{fig:blob_frag}
	\end{figure*}

After the initial compression, the magnetic field overexpands back to eventually reach an equilibrium close to the initial condition. The blobs are then stretched out vertically by the expansion of the magnetic field. The final result is that the blobs spread out into a highly fragmented flow (Fig.~\ref{fig:maps_rm} and Fig.~\ref{fig:blob_frag}).

This effect can be explained as follows: each blob falls compressing the field lines ahead of it. At the same time it drags the field lines that cross it. This field line shifting is not uniform, because of the shape of the blob and of the presence of other nearby blobs, and creates a differential stress back on the blobs. This stress becomes chaotic with the back-expansion of the field. Thus, the expansion of the field acts as a mixer that fragments the flow.

The field line shifting also perturbs the footpoint of the magnetic channel, leading to a chaotic change of its section close to the chromosphere. These changes produce pressure gradients that trigger spicule-like upflows, not investigated in this paper.
 
After $t \sim 100$ s the expansion of the field ends and the initial magnetic field configuration is fully restored. The fragments continue to flow along the field lines with only a fraction of the initial velocity  ($ \sim 100$ km/s). The cool and dense plasma appearing after the blobs are channelled (Fig.~\ref{fig:maps_rm} and movies 2a and 2b) is not caused by thermal instability, as in \cite{Fangetal2015,Xiaetal2014,Xiaetal2016}. Instead, this is part of the fragments that, while falling and distorting, occasionally crosses the 2D plane of the figure (and of the movies).

\begin{figure*}[!t]
	\centering
	\includegraphics[width=12cm]{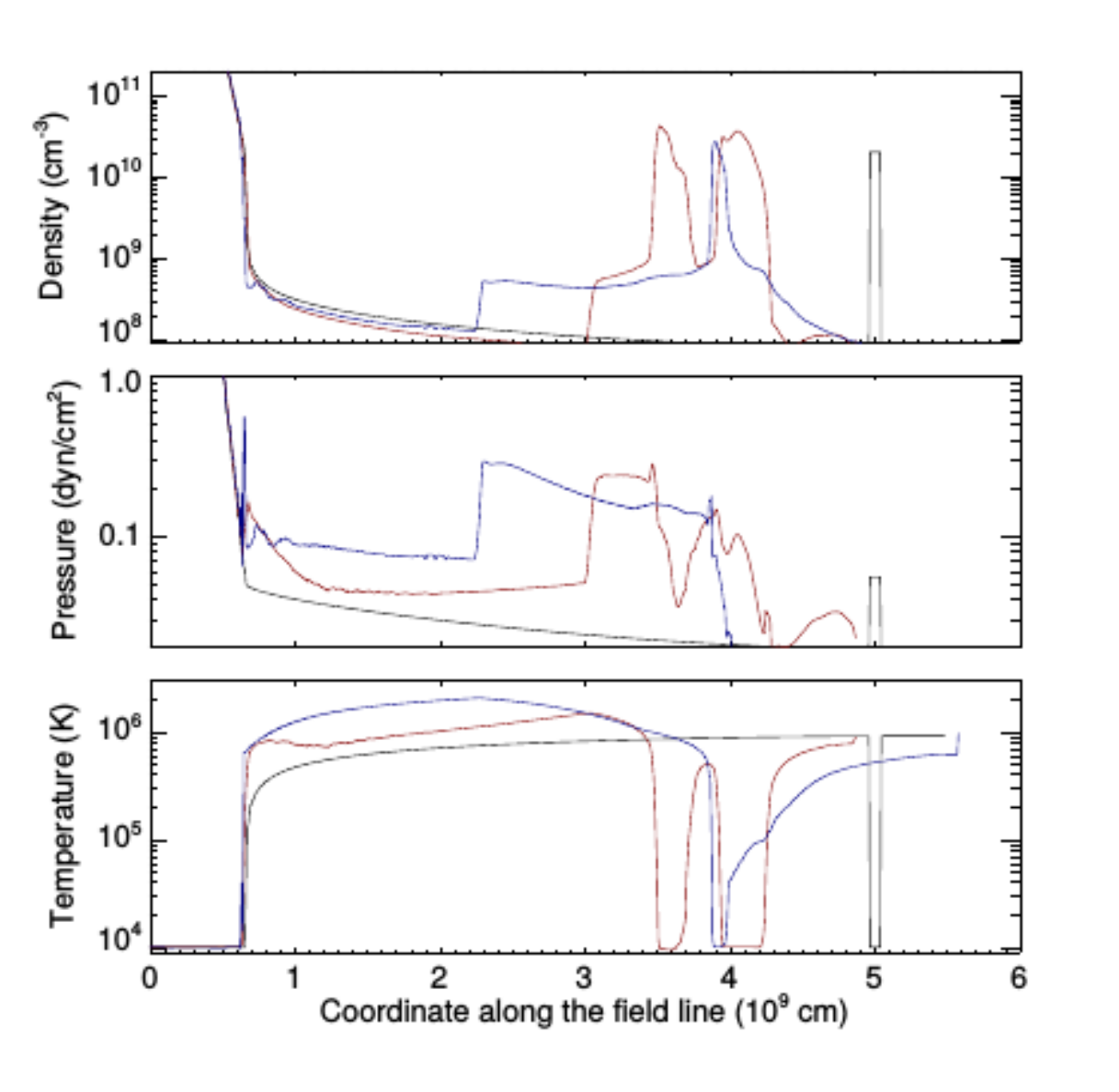}
	\caption{Profiles of density, temperature and thermal pressure along a magnetic field line that encounters one of the blobs at the initial time. The profiles are taken at time t = 0 s (black line), 50 s (red) and 100 s (blue). The chromosphere is at the left-hand side. }
	\label{fig:Magcomp}
	\end{figure*}

The initial velocity of the blobs, $300$ km/s, is greater than the sound speed in the corona ($c_s$):

\begin{equation}
c_s = \sqrt{\frac{\gamma p}{\rho}} = \sqrt{2 \gamma n_H k_B T} \sim 150 ~ km/s
\end{equation}

\noindent
where we conservatively assume adiabatic shocks. Therefore, shocks propagate ahead of the blobs.
These shocks compress and heat the coronal material between the blobs and the  chromosphere to temperatures  $T \sim 1-2 \times 10^6$ K. An estimate of the shock speed ($v_{sh}$) can be evaluated considering that they move along a path $\delta{l}\sim 4.3 \times 10^4$ km until they hit the chromosphere for a time $\delta{t} \sim 140$ s:

\begin{equation}
	v_{sh} = \frac{\delta{l}}{\delta{t}} =  \frac{4.3\times10^4}{140} \sim 300 ~~ km/s
\end{equation}

\noindent
or by the Rankine-Hugoniot formula for isothermal shocks:

\begin{equation}
	\frac{\rho_2}{\rho_1} = \frac{v_1^2}{c_s^2},\quad v_1  \sim 300 ~ km/s
\end{equation}

\noindent
where $\rho_2$ and $\rho_1$ are the density of post- and pre-shock medium, respectively (Fig.~\ref{fig:Magcomp}), $c_s$ is the sound speed, and $v_1$ is the velocity of the pre-shock medium in the reference frame of reference the shock (so it is the velocity of the shock in our reference frame). The shocks stream along the field lines, because the Alfven speed ($v_A = B / \sqrt{4 \pi \rho } \sim 1000$ km/s) is much higher than the shock speed $v_{sh}$ \citep[e.g.,][, p.184]{Priest2014}. These shocks are slow-mode shocks. In the vertically descending phase, the blobs are compressed in the direction perpendicular to local magnetic field. The compression and solar gravity together accelerate the leftward expansion of the blobs along the magnetic field lines, which generates and drives the slow-mode shocks.

For a more quantitative grasp of the structure and dynamics of the fragments, Fig.~\ref{fig:Magcomp} shows plots of the density, pressure and temperature all along a magnetic field line that crosses one of the blobs at the initial position, at 3 different times. The blob is a squared bump in the density and pressure, and a dip in the temperature at time $t = 0$. At t=50~s, the blob has moved leftwards along the line by $\sim 10^4$ km, while being highly spread out and deformed. The shock front is clearly visible ahead of it in the pressure and density, much less in the temperature due to the fact that thermal fronts in corona propagates faster than shocks, so they are almost isothermal \citep[the relative speed of these two phenomena can be triggered by changing the factor $\phi$ in the saturated thermal conduction,][]{Orletal2005}.  At $t = 100$~s the front density peak has declined while  the central peak has not moved much leftwards. The reason for this apparent rest is that the field line has significantly stretched between 50 and 100 s (by $\sim 20$\%), because of the magnetic field back-expansion. So the distance from the chromosphere at t=100 s should be scaled as well.  On the other hand, in spite of this effect, the shock front has moved leftward, by $10^4$ km. The temperature has clearly increased to a peak of $\sim 2$ MK. 

From the velocity of the shocks one can derive their ram pressure. Considering that the post-shock plasma is at density $n_{sh} \sim 6 \times 10^8 cm^{-3}$:

\begin{equation}
	p_{ram\_sh} = n_{sh} \mu m_p v_{sh}^2  \sim 1 ~~ dyn ~ cm^{-2}
\end{equation}

\noindent
their ram pressure is much lower than the ambient magnetic pressure ($p_m~\sim~10$ dyn cm$^{-2}$ at the blobs initial position).

\subsection{Dense model (DM)}

\begin{figure*}[!t]
	\centering
	\includegraphics[width = 14cm]{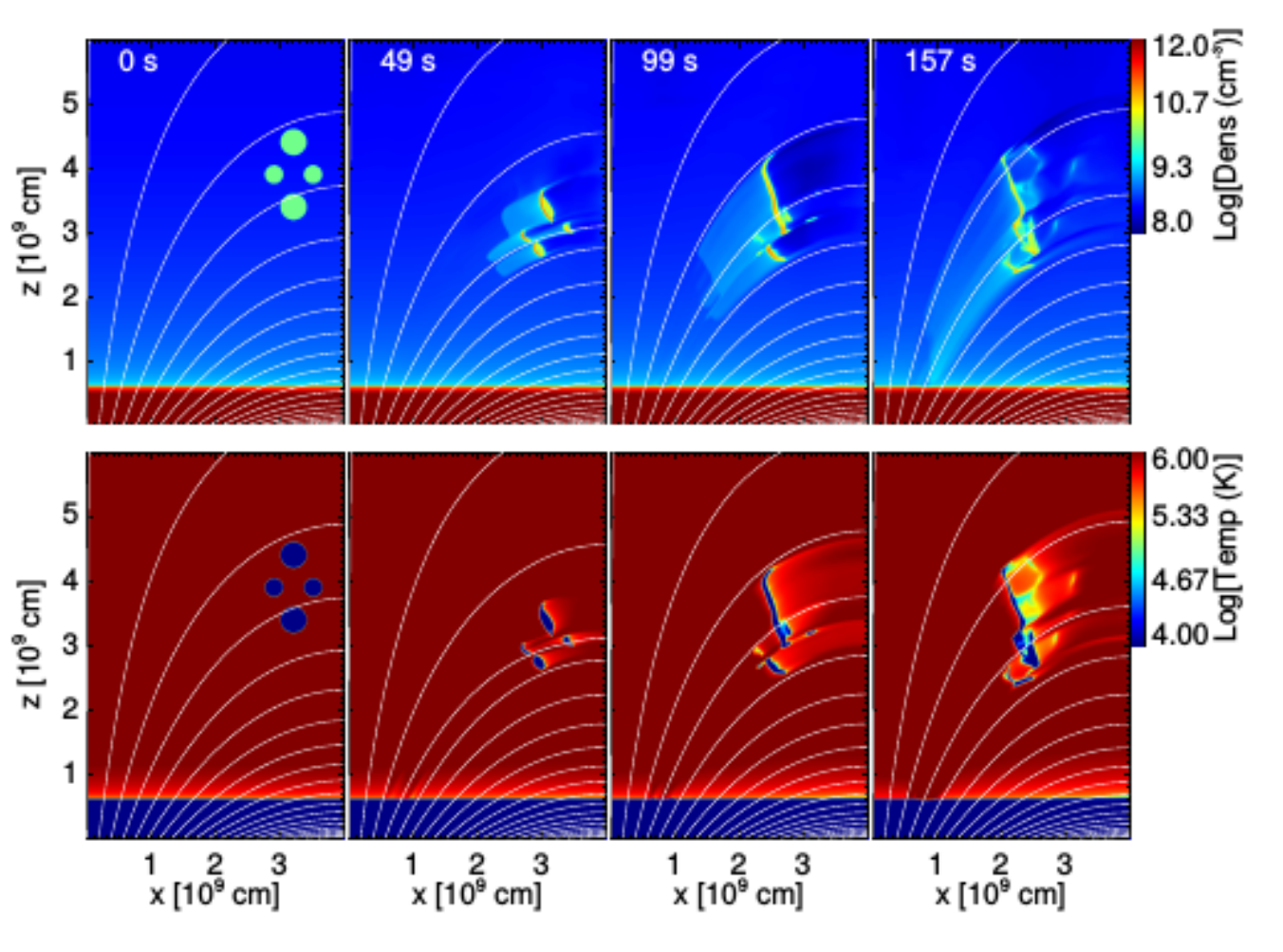}
	\caption{Same as Fig.~\ref{fig:maps_rm} but for the Dense Model (DM).}
	\label{fig:maps_dm}
	\end{figure*}

In the DM model the ambient medium is denser and the blobs have half the density as in the previous case. As a consequence of the lower density contrast, the blobs are channelled sooner along the magnetic field lines (Fig.~\ref{fig:maps_dm}), because the magnetic pressure has to balance a lower ram pressure than in the RM. The denser atmosphere affects the shape and the velocity of the blobs. In this case, the velocity is $v \sim 70$ km/s (to be compared with $\sim100$ km/s in the RM), and the blobs appear to be thin slabs at higher density (n $\sim 10^{11}$ cm$^{-3}$). The atmosphere conditions affect also the ratio between post- and pre-shock region that leads to a velocity of the shocks of $250$ km/s. Moreover the blob motion is much slower and the shocks are much fainter than in the reference model.

\subsection{Cold model (CM)}

\begin{figure*}[!t]
	\centering
	\includegraphics[width = 14cm]{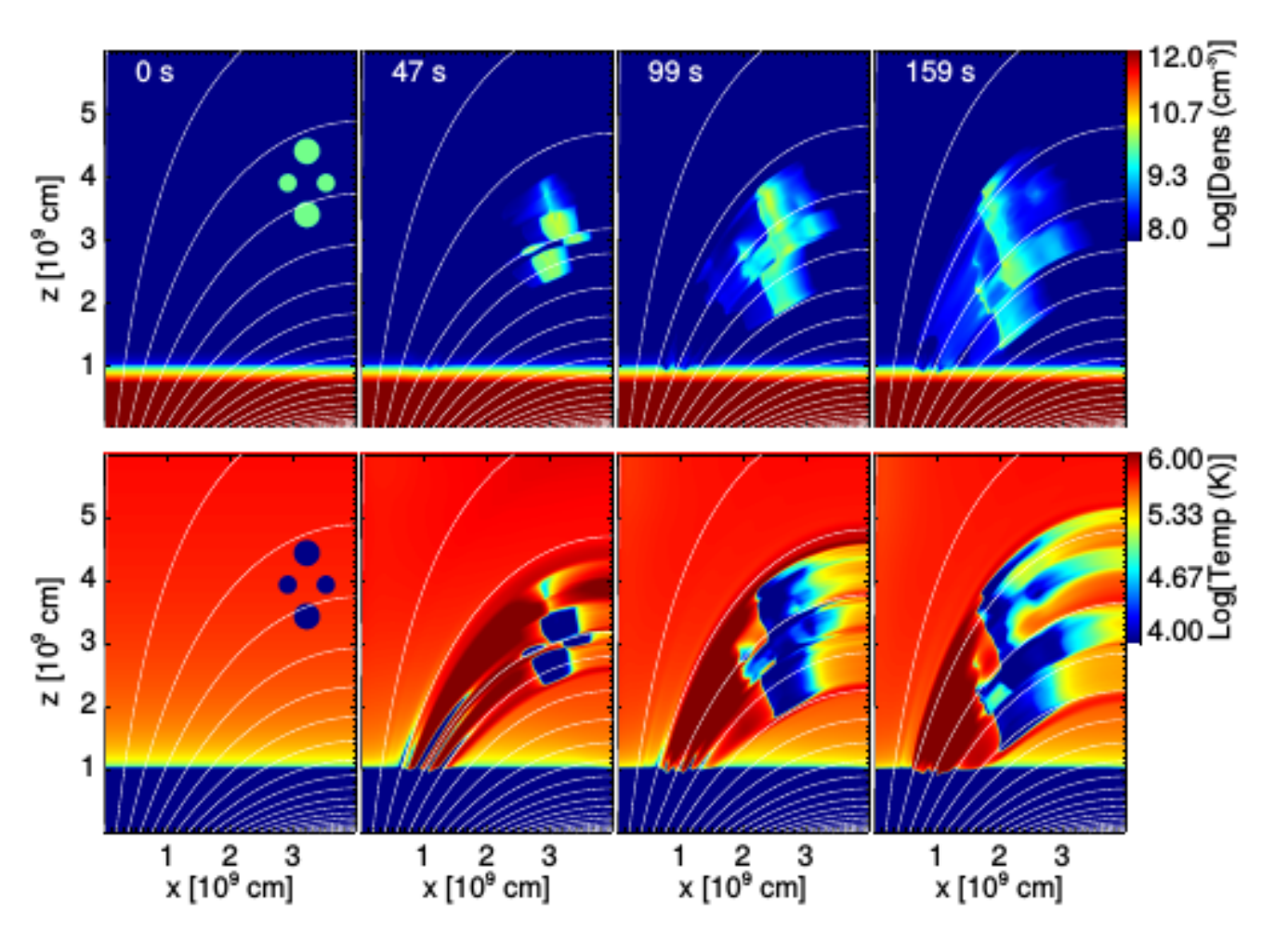}
	\caption{Same as Fig.~\ref{fig:maps_rm} but for the Cold Model-CM.}
	\label{fig:maps_cm}
	\end{figure*}

Overall, in this case the blobs fragment less than in the previous ones. While they move toward the chromosphere, the blobs are still highly deformed but not squashed into slabs, essentially because the initial density ratio between the blobs and ambient medium is a factor 20 higher than in DM (Fig.~\ref{fig:maps_cm}). The effect of the dynamics on the magnetic field is the same as in the previous cases, because DM and CM share the same magnetic field configuration and blobs density, so Eq.~(\ref{eq:prescomp}) still holds. The change in the atmosphere instead affects the residual velocity of the blobs after the magnetic field expansion which in this case is $120$ km/s, and the velocity of the shock, which is $180$ km/s.

\subsection{Weak field model (WM)}

\begin{figure*}[!t]
	\centering
	\includegraphics[width = 14cm]{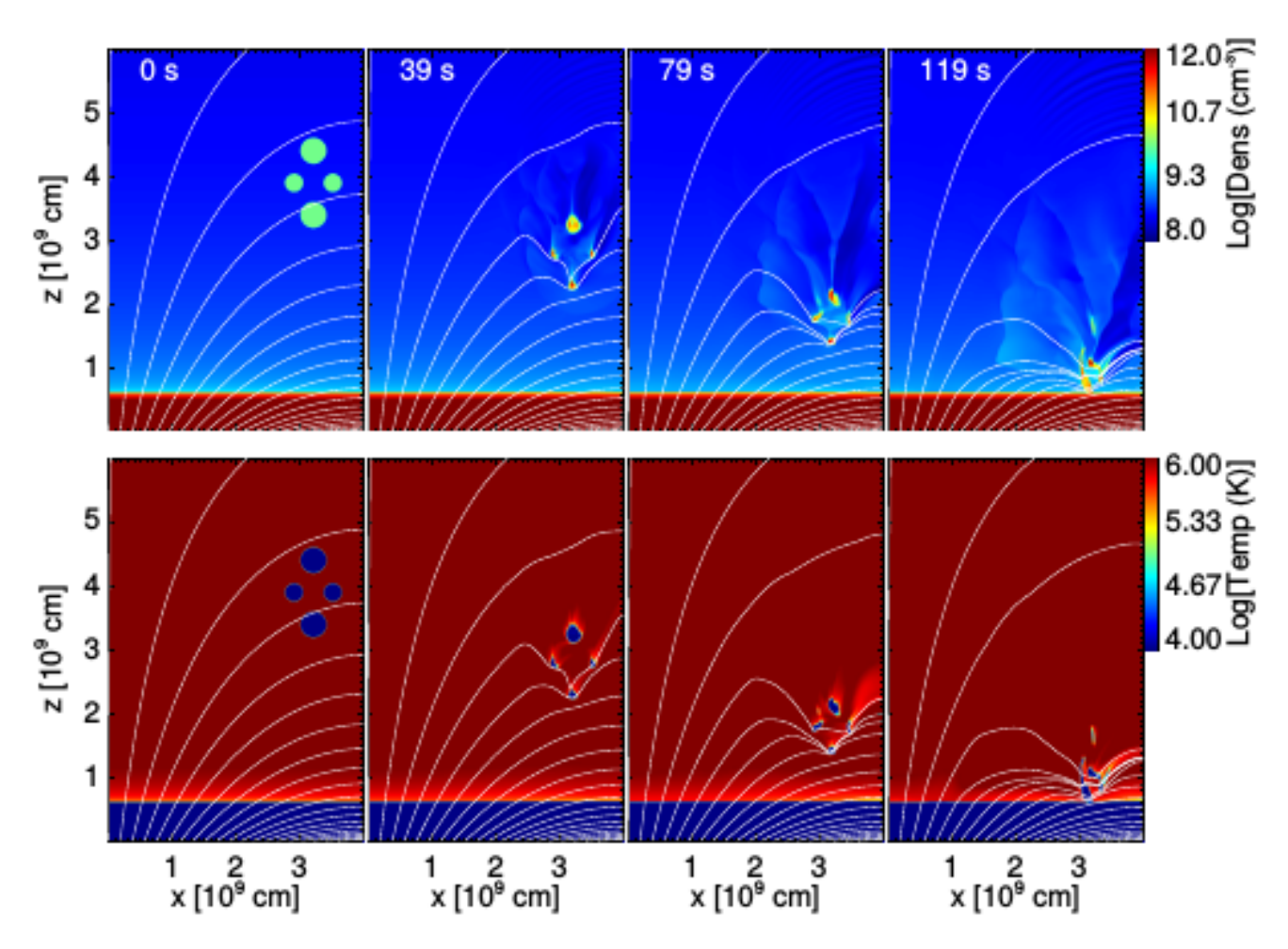}
	\caption{Same as Fig.~\ref{fig:maps_rm} but for the Weak-field Model-WM.}
	\label{fig:maps_wm}
	\end{figure*}

In the WM simulation, the magnetic pressure never balances the ram pressure of the blobs, even if it increases because of the falling blobs similarly to the previous cases. Therefore, Eq.~(\ref{eq:prescomp}) holds at all times and the blobs fall vertically until they impact the chromosphere, as shown in Fig.\ref{fig:maps_wm}. The shock ahead of the blobs is very weak and rapidly damped by the compressed magnetic field that envelopes the blobs during the falling. This simulation is quite similar to those shown in \cite{Reaetal2013}, but quite different from the observed evolution. Therefore, it puts a lower limit to the ambient magnetic field intensity ($\sim1$ G), but we will no longer discuss it in the following.

\subsection{Synthetic emission}
\label{sec:s_emis}

To compare the results of the simulations with the observation, from the output of the simulations we have synthesized the emission in the AIA 171 \AA~ channel. The filterband of this channel includes a strong Fe IX line with a temperature of maximum formation of $\sim 10^6$ K.
We have calculated the emission in each cell of our 3D computational domain as

\begin{equation}
I_{171}(x,y,z) = G_{171}[T(x,y,z)] n_e^2(x,y,z)
\end{equation}

\noindent
where $G_{171}$ is the response of the channel as a function of the temperature of the emitting plasma (available from the SolarSoftware package).
Then, we have integrated $I_{171}(x,y,z)$ along two possible line of sights, i.e., along Z and along Y.
To account for absorption from optically thick plasma, we have neglected the emission from cells with a density greater than $10^{10} cm^{-3}$ \citep{Reaetal2013} and beyond, along the line of sight.

For the DM simulation, the background atmosphere is at the same time relatively dense and hot and it fills the whole computational domain, which is much larger than the volume involved in the dynamics driven by the falling fragments. For this reason, the atmosphere is very luminous in the selected AIA channel, and, when we integrate it along the line of sight, it dominates over the emission excess produced by the fragments. Since in the observation the volume of the background atmosphere is not so large and its emission is not important for our analysis, only for this case we have decided to integrate only the emission through the flux-tube in which the blobs propagate.

\begin{figure*}[!t]
	\centering
	\includegraphics[width=12cm]{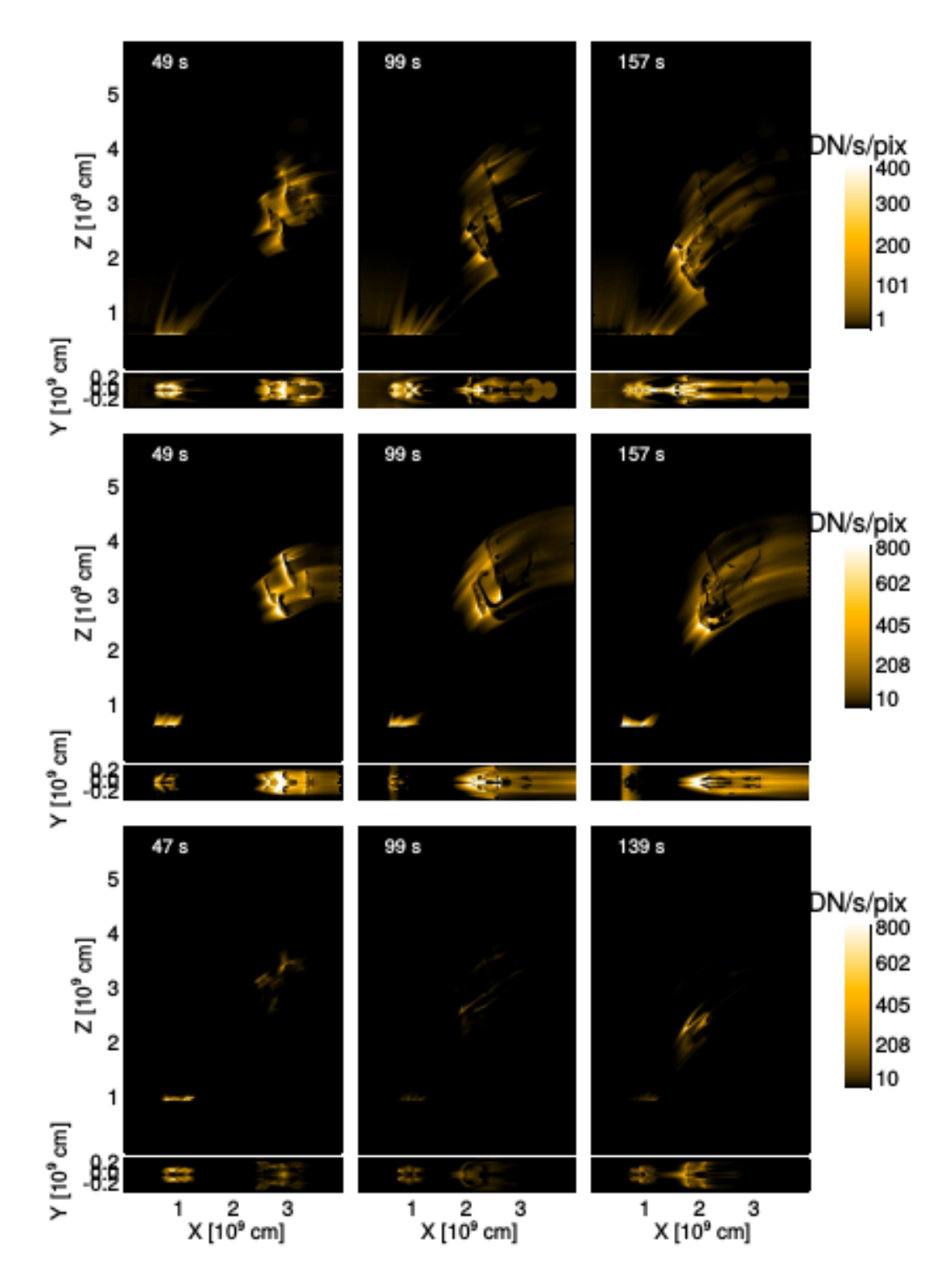}
	\caption{Images of the integrated emission (square root color scale) in the AIA 171 \AA~  channel, at the labelled times from top to bottom, respectively, for RM, DM, CM. Each panel includes both the map integrated along Y (top) and along Z (bottom). (See Supplementary Movie 4)}
	\label{fig:171syntmaps}
	\end{figure*}

Fig.~\ref{fig:171syntmaps} shows maps of the expected emission along both lines of sight, in a form similar to the observed ones shown in Fig.~\ref{fig:iniobs}b.

In all the models, as the blobs move, bright fronts develop ahead of them. Initially, they  are thin shells just in front of the blobs but later they extend much beyond them, in the form of elongated filaments.

\begin{figure*}[!t]
	\centering
	\includegraphics[width=10cm]{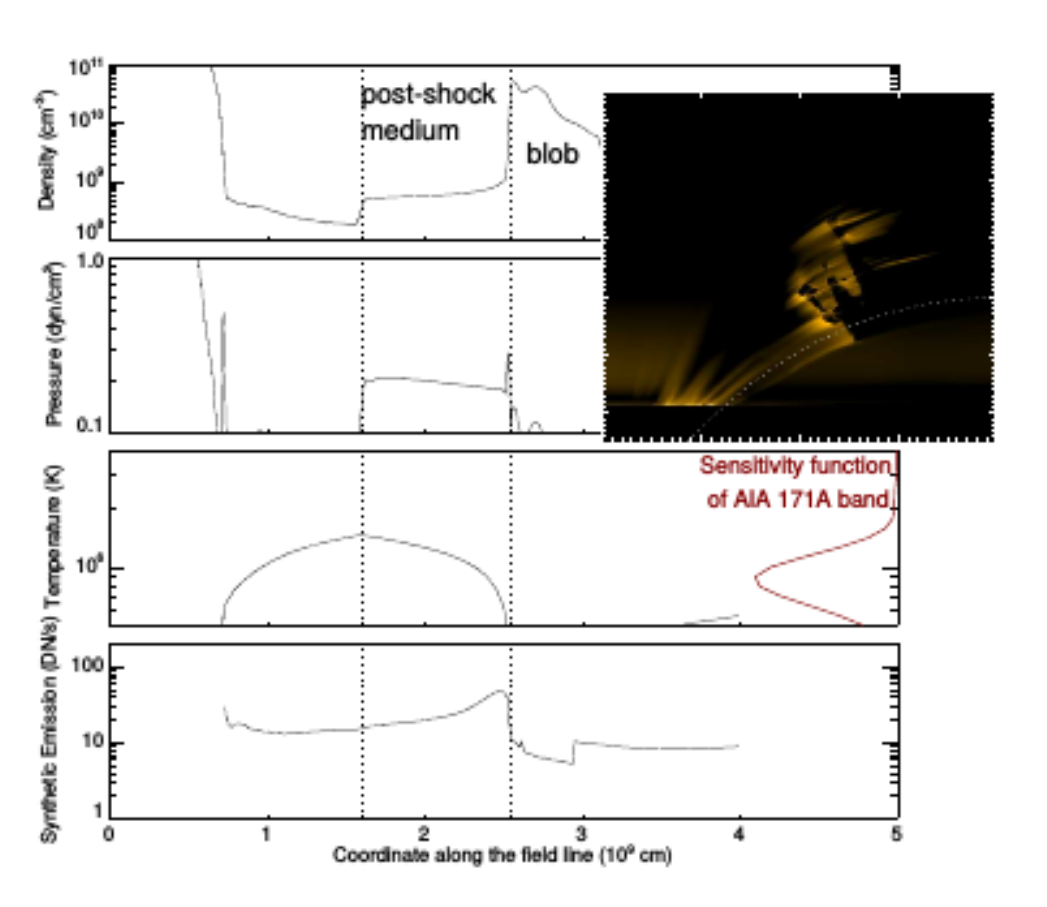}
	\caption{Density, temperature, pressure and integrated emission along the white dashed line shown in the side image (which is taken at t= 100 s). The sensitivity function of the AIA 171 \AA~ filter band is plotted (red line) on the right side of the temperature plot. The dotted vertical lines enclose the post-shock region.}
	\label{fig:compdynemis_mid}
	\end{figure*}

To investigate where the emission comes from, in Fig.~\ref{fig:compdynemis_mid} we have related the evolution of the brightening to the dynamics by plotting density, pressure, temperature and synthetic integrated (Y-line of sight) along a magnetic field line, at time t = 100 s, for the reference model (RM). We have selected a magnetic field line that intersects only one blob.

As shown in Fig.\ref{fig:compdynemis_mid}, the brightest emission comes from the post-shock medium ahead of the blobs. However, the high emission extends beyond the shock front down almost to the top of the chromosphere. The shock front is visible in the density profile, which is purely along the field line, but much less in the emission profile, which is integrated along the line of sight. The shock fronts are not aligned to the line of sight and therefore, the integration along that line washes out any sharp front. The emission in the unperturbed medium beyond the shock is higher than initially. The reason is that the shock heats the medium in which it propagates. Therefore, a thermal front also develops, and it moves downwards along the field line by pure thermal conduction. Even in the presence of saturation, we can estimate (e.g. \citealt{Reaetal2014}) that the conduction time scale over a length of $\sim 10^4$ km, a temperature $\geq 1$ MK and a density $\sim 10^8$ cm$^{-3}$ is below 100 s, in agreement with the simulation result. As a consequence, the temperature rises up also below the shock to values in the range of higher sensitivity of the 171 \AA~ channel. The emission in this channel therefore increases. Similar effects occur also in the other simulations (DM, CM), although with some quantitative differences.

\begin{figure*}[!t]
	\centering
	\includegraphics[width=16cm]{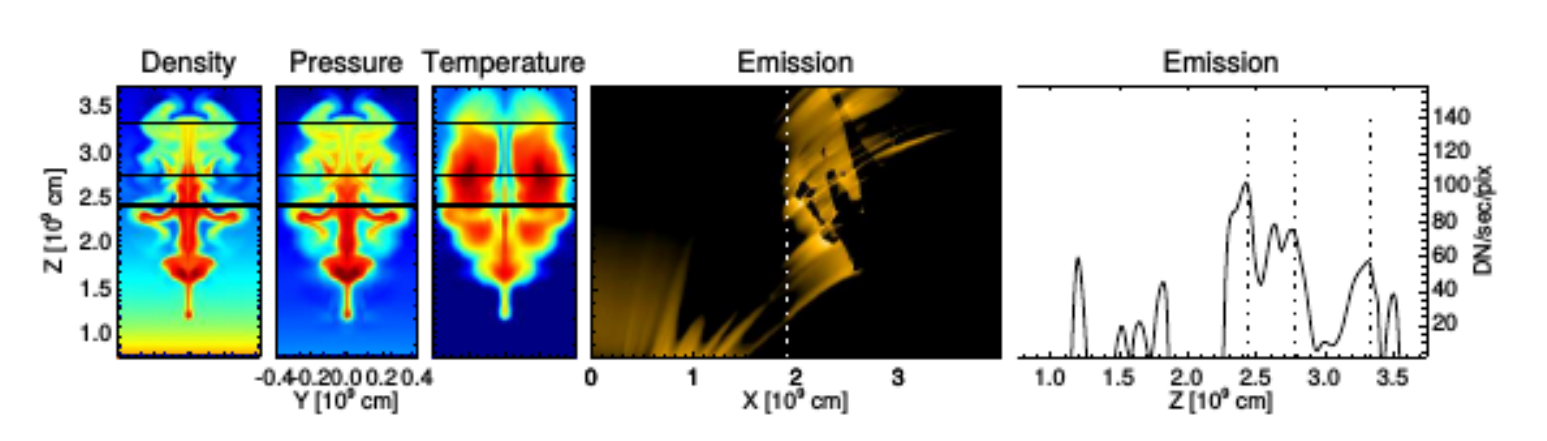}
	\caption{Density, pressure and temperature (left 3 images) for the RM at time t = 100 s, in the plane perpendicular to that of the integrated AIA 171 \AA~ emission (middle) passing through the dotted line, and plot of the integrated AIA 171 \AA~ emission along the same line (right). The black lines on the left panel mark the position of the peaks in the plot on the right (dotted lines).}
	\label{fig:filcomp}
	\end{figure*}

Another interesting issue is the fact that the emission is finely structured into bright fibrils. To understand why, Fig.~\ref{fig:filcomp} compares the maps of the emission with transversal maps (YZ) of density and temperature at an X position across the filaments, i.e. across the post-shock region in front of the blobs.

\begin{figure*}[!t]
	\centering
	\includegraphics[width=8cm]{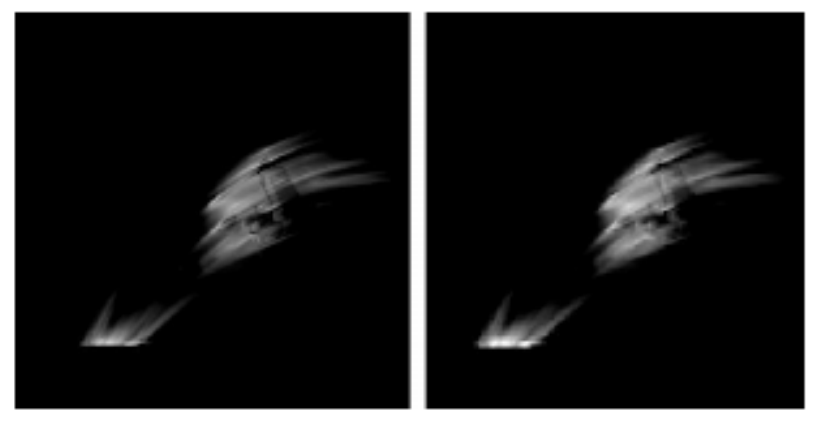}
	\caption{Images of the integrated emission in the AIA 171 \AA~ channel for the CM at the full resolution of the simulation (left) and at the AIA resolution (right), at time t = 50 s.}
	\label{fig:comp_res_cold}
	\end{figure*}

The transversal maps show that the shocked medium is highly sub-structured inside the magnetic channel. For a continuous flow propagating along the field we would expect a well defined shock-front propagating ahead of it, and therefore an emission uniformly increasing along the channel. Instead, we have four blobs that move initially not aligned to the magnetic field. As such they are able to mix the magnetic field lines during the initial phase of the evolution, and the feedback from the field is chaotic,
leading to a further fragmentation similar to those typical of hydrodynamic instabilities. However, when the spatial resolution of AIA is taken into account (Fig.~\ref{fig:comp_res_cold}), these fine filaments are blurred into much thicker ones, as observed. 

\begin{figure*}[!t]
	\centering
	\includegraphics[width=10cm]{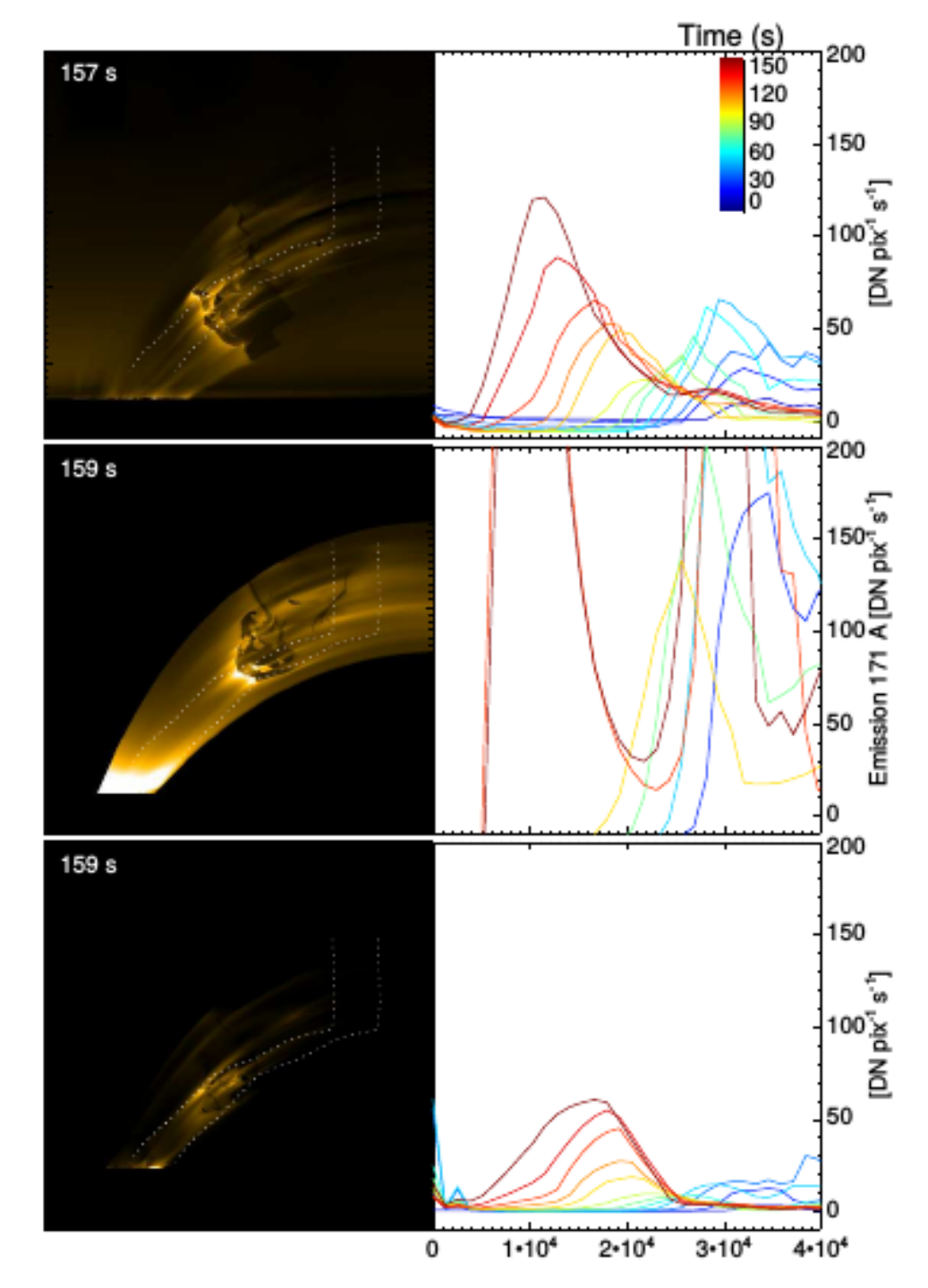}
	\caption{{\it Left}: Integrated emission images as in Fig.\ref{fig:171syntmaps}(top) for RM (top), DM (middle) and CM (bottom). Inside the strips (dotted lines) we compute emission profiles for comparison with the observed ones in Fig.~\ref{fig:blobtraj1}c. {\it Right}: Emission profiles along the strips marked in the left panels, with the same format and resolution as in the observed ones shown in Fig.~\ref{fig:blobtraj1}c. The profiles are sampled at intervals of 12 s for the RM and CM, and 24 s for the DM (for the sake of clarity), at progressing times (from blue to red) to t = 150 s. The width of the strip is comparable to that in Fig.~\ref{fig:blobtraj1}a.}
	\label{fig:comp_emis}
	\end{figure*}

Finally, in Fig.~\ref{fig:comp_emis} we show synthetic emission profiles from simulations RM, DM and CM to be compared with the observed ones in Fig.~\ref{fig:blobtraj1}c. We remark that our simulations describe only the phase in which the blobs brighten the magnetic channel, which corresponds to the time and space between the dashed lines in Fig.~\ref{fig:blobtraj1}b. As we did for the observation, we extract emission profiles along a strip, with the same width as the one marked in Fig.~\ref{fig:blobtraj1}a.  As we did for the observation, we subtracted the emission of the initial frame. 
	
The height and shape of the profiles are different from one case to the other.  
In the DM the temperature of the post-shock region exceeds the range in which the 171 \AA~ channel is sensitive, and only a small fraction of the post-shock region near the blobs is able to emit efficiently in that band, leading to an emission with many spikes, which largely exceed the observed count rate. Instead, both in RM and CM, a larger fraction of the post-shock region emits in the channel and the emission appears to be smoother. In both cases we also clearly see bright fronts moving to the left, i.e. to the solar surface and a growing emission when approaching the surface.
The different intensity of the emission, instead, is related to the different ambient density where the shocks propagate. Overall, we find that the RM case has the best agreement with the observations, and is able to reproduce both the growing emission peak toward the end of the path and a similar DN rate.

\section{Discussion and Conclusions}

We studied the downfall of blobs of plasma channelled by the magnetic field toward an active region. These blobs were erupted by a M-class flare event on 7 June 2011, and showed a ballistic motion while still far from the active region. We see the blobs in absorption and we constrain their density to be around $1 - 2 \times 10^{10}$ cm$^{-3}$, according to the method in \cite{LaneRea2013}. As the interaction with the magnetic field becomes strong, they are deviated from their trajectory and channelled by a magnetic flux channel. During the channelling, the flux tube brightens in the 171 \AA~ EUV channel of the AIA instrument, and the blobs disappear. We investigated the channelling process with the aim to explain the brightening of the magnetic channel.

We considered a model of a magnetized atmosphere with a curved topology of the magnetic field and a complete solar atmosphere from the chromosphere to the corona, and included all the physical terms of interest, in particular, gravity, radiative losses, thermal conduction along the field lines, and magnetic induction. The model solved numerically the magneto-hydrodynamic equations in 3D Cartesian geometry, implemented in the PLUTO parallel code. The blobs are modelled as spheres with a downward velocity of $300$ km/s not aligned with the magnetic field, different radii ($1.4-2 \times 10^8$ cm), density ($1-2 \times 10^{10}$ cm$^{-3}$), and a temperature of $10^4$ K. We tested the role of the atmosphere as well as of the magnetic field by exploring ambient atmosphere with three different ambient densities and two different magnetic field intensities, with the same topology.

The blobs started to fall vertically in all the models, but only in the case in which the magnetic field is strong (170G in transition region and 15G at initial blobs position) they are channelled and deviated from their trajectory. In the case of weak magnetic field, the blobs simply fall without any deviation, similar to \cite{Reaetal2013}. This case is far from our target evolution and provides a lower limit to the conditions of the ambient magnetic field. 

The initial velocity of the blobs largely exceed the ambient sound speed, so shocks are generated. The behaviour of these shocks depends on the physical condition of the model explored, but with a common dynamics: they propagate ahead of the blobs inside the magnetic flux tube, in which the blobs are channelled, along the field lines. 

Another effect of the dynamics is that the blobs are strongly deformed, even further fragmented, during their motion. Two factors contribute to this effect: a) the field lines are chaotically displaced downwards and then back upwards, thus being mixed and determining braiding and a differential stress on the blobs, b) the blobs are squashed in the direction of motion. The former effect is common to all the confined models, while the latter depends strongly on the density and pressure of the ambient atmosphere, the larger the density (pressure) the stronger is the compression, and it is also affected by the compression of the magnetic field lines in the initial stage of the evolution. 

By synthesizing the emission in the 171 \AA~ EUV band, we identified the post-shock region as the main source of the brightening ahead of the blobs. The emission depends on the density and temperature of the ambient atmosphere. The former heavily influences the intensity, because of the dependence on the square of the density, the latter acts more on the shape and size of the emission because of the narrow temperature range of channel sensitivity.  As a consequence, for the simulation with high ambient density the intensity of the emission produced is too high and its profile along the field lines does not match what we observe. Instead, for the other two densities, the cooler and tenuous atmospheres give a  shape and intensity that better agrees with the observations, best for the one that we called the reference model (RM). 

The simulations show that the emission comes not only from the the post-shock region, but the whole magnetic channel between the blobs and the chromosphere is activated, well before the shock arrives at the chromosphere. The reason is that the shock compresses and heats the medium it crosses, and the heat front propagates downwards faster than the shock, making the unperturbed medium enter more in the AIA sensitivity range, with this assumption for $\phi$ for coronal condition \citep{Orletal2005}. This makes the emission contrast between the pre-shock and post-shock medium lower. More important, the shock are ultimately never visible as well-defined fronts in our scenario, for another reason. Each fragment or blob produces its own shock front, and, since the blobs are different and not aligned along the line of sight, also the shock are misaligned in time and space and washed out along the line of sight. 

Overall, our reference model provides the best match with the evolution of the channelled fragment that we selected in the observation. The parameters, i.e. size and density, of the blobs that we assumed in this simulation are well within the constraints provided by the data analysis. Therefore, we obtain a self-consistent scenario. Moreover, within our limited exploration of space of the parameters, our modeling provides us with constraints on, and therefore probes, the ambient medium, and in particular on the ambient coronal magnetic field ($\sim 10$ G) and density ($\sim 10^8$ cm$^{-3}$). 

Several general considerations descend from this study. We find that falling fragments are disrupted because of the chaotic interaction with a strong ambient magnetic field. The disruption occurs just when the fragments are deviated and channelled by the field. The ram pressure of the fragments displaces and compresses differentially the field lines, which react back and shuffle the blobs. Therefore, a misalignment of dense falling plasma with the local more intense magnetic field lead to a disruption of falling clouds. This evolution also becomes a signature of a strong local perturbation of the magnetic field. We considered simplified spherical blobs with homogeneous density, but in the reality they can be highly inhomogeneous, thus making the mixing and fragmentation even more chaotic.  

Indeed analogous effects are expected even for more continuous streams, instead of a set of blobs. The magnetic field lines can still be chaotically displaced downwards and the plasma can be squashed in the direction of the motion, depending on the magnetic field topology and on the trajectory of the stream. As a result, an initially continuous stream can be fragmented or become density-structured after interacting with the curved magnetic structures. This fact can be relevant for the accretion on young stars, which occurs along magnetic flux tubes between the disk and the stellar surface. It is plausible that even if an accretion stream is almost continuous in proximity of the disk, it may interact  with the more complex magnetic field topology of the stellar corona when approaching the star surface, thus experimenting the effects
described above. We may expect, therefore, that accretion streams in young stars may be commonly density-structured at impact regions.

Regarding the emission, this study shows another mechanism that lead to an excess of emission in high energy bands. 1D/2D accretion models show higher emission due to a stationary shock produced by a continuous accretion column on the stellar surface \citep{Saccetal2010,Orletal2010,Orletal2013}. \cite{Reaetal2013,Reaetal2014} show that the impact of massive but isolated fragments also lead to hot brightenings. Here we show that falling fragments eventually channelled by the magnetic field do not brighten themselves, rather they activate the channel and make it bright, because of shock propagation and heating. This early further fragmentation and activation of the magnetic channel is certainly a considerable difference from the evolution studied in fragments that do not interact so strongly with the magnetic field. As described in \cite{Reaetal2013,Reaetal2014}, in that case the disruption of the fragments and the brightening are due exclusively to the impact on the dense chromosphere. One important implication for stellar accretion is that we might have emission excess also if the accretion flow interacts with a coronal magnetic field that is not strictly aligned to the flow, e.g. with a solar-like corona with intense active regions.

In this study, we have assumed that the initial vertical trajectory of blobs is on the symmetry vertical plane of the magnetic field. We expect possible important effects in cases in which the falling fragments are ``out of axis" and stress the magnetic field in a non-symmetric way. This is to be investigated in a future study.

\acknowledgements{AP, FR, and SO acknowledge support from the Italian \emph{ Ministero dell'Universit\`a e Ricerca}. P.T. was supported by contract SP02H1701R from Lockheed-Martin to the Smithsonian Astrophysical Observatory, and by NASA grant NNX15AF50G. AP, FR and PT acknowledge ISSI for support to the team "New diagnostics of particle acceleration in solar coronal nanoflares from chromospheric observations and modeling”. PLUTO is developed at the Turin Astronomical Observatory in collaboration with the Department of Physics of the Turin University. We acknowledge the HPC facilities SCAN, of the INAF - Osservatorio Astronomico di Palermo and the CINECA Award HP10CCVR17, for the availability of high performance computing resources and support. VisIt \citep{HPV:VisIt} (Fig.~\ref{fig:blob_frag} and Movie3) is supported by the U.S. Department of Energy with funding from the Advanced Simulation and Computing Program and the Scientific Discovery through Advanced Computing Program. CHIANTI is a collaborative project involving the NRL (USA), the Universities of Florence (Italy) and Cambridge (UK), and George Mason University (USA).} 

\bibliographystyle{aa}
\bibliography{blob3d}
\end{document}